\documentclass[apj]{emulateapj}
\slugcomment{Accepted for publication in ApJ}
\newcommand\xmm{{\it XMM-Newton }}
\newcommand\chandra{{\it Chandra}}
\newcommand\suzaku{{\it Suzaku }}

\newcommand\kev{{\rm~keV}}

\newcommand\ep{EPIC-pn}
\usepackage{graphicx}
\usepackage{amsmath}
\usepackage{apjfonts}


\def\***#1{{\sc #1}}
\def\plan#1{\relax}
\def\Plan#1{\relax}
\def\PLAN#1{\relax}

\def\lta{\mathrel{\spose{\lower 3pt\hbox{$\mathchar"218$}}
     \raise 2.0pt\hbox{$\mathchar"13C$}}}
\def\gta{\mathrel{\spose{\lower 3pt\hbox{$\mathchar"218$}}
     \raise 2.0pt\hbox{$\mathchar"13E$}}}

\def\mathnew{\mathsurround=0pt}

\def\simov#1#2{\lower .5pt\vbox{\baselineskip0pt \lineskip-.5pt
\ialign{$\mathnew#1\hfil##\hfil$\crcr#2\crcr\sim\crcr}}}

\def\simless{\mathrel{\mathpalette\simov <}}

\shorttitle{Spectral variability of IRAS 18325-5926} \shortauthors{Tripathi et al. }

\begin{document}

\title{Spectral variability of IRAS 18325-5926 and constraints on the
  geometry of the scattering medium.}

\author{Shruti Tripathi\altaffilmark{1}, R. Misra\altaffilmark{1}, G. C. Dewangan\altaffilmark{1}, J. Cheeran\altaffilmark{2}, S. Abraham\altaffilmark{2} and
  N. S. Philip\altaffilmark{2}}
\altaffiltext{1}{Inter-University Center for Astronomy and
  Astrophysics, Ganeshkhind, Pune-411007, India;
  stripathi@iucaa.ernet.in}
\altaffiltext{2}{Department of Physics, St. Thomas College,
  Kozhencherri-689641, India}

\begin{abstract}
  We analyze Suzaku and XMM-Newton data of the highly variable Seyfert
  2, IRAS 18325-5926. The spectra of the source are well modeled as a
  primary component described as an absorbed power law and a secondary
  power law component which is consistent with being scattered
  emission from an on-axis extended highly ionized medium. We show
  that while the primary component varies on a wide range of
  timescales from $10^{4} - 10^{8}$ s, the scattered emission is
  variable only on timescales longer than $10^{5}$ s. This
  implies that the extent of the scattering medium is greater than
  $10^{16}$ cm. The ratio of the scattered to primary flux ($\sim
  0.03$) implies a column density for the scattering medium to be
  $\sim 10^{23}$ cm$^{-2}$. We argue that for such a medium to be
  highly ionized it must be located less than $10^{17}$ cm from the
  X-ray source. Thus we localize the position and extent of scattering
  region to be $\sim$ a few $\times 10^{16}$ cm, with an average
  particle density of $\sim 10^{6}$ cm$^{-3}$. We consider the
  physical interpretation of these results and as an aside, we confirm
  the presence of a broad Iron line emission in both the {\it XMM-Newton}
  and {\it Suzaku} observations.
\end{abstract}

\section{Introduction}
The Unified Model for Seyfert galaxies predicts that the differences
observed between type 1 and type 2 Seyfert galaxies are primarily due
to orientation effects \citep{Anto93}. According to this model, there
is a geometrically thick torus surrounding the accretion disk around
the central supermassive black hole. Type 1 Seyfert galaxies are those
active galactic nuclei (AGN) seen at low inclination angles (i.e. nearly face on) which provide an unobstructed view of the inner central region. Broad optical emission lines, which originate close to the black holes, are observed
along with the narrow ones, originating further from the source. In
the X-ray regime, the spectra is typically a power law one (showing
sometimes a soft excess) and affected mostly by absorption in our own
Galaxy. In contrast, type 2 Seyfert are those AGN which are observed
at high inclination angles (i.e. nearly edge on) and hence their
central engine is obscured by the dusty torus. The broad optical lines
are not observed here but the narrow ones which originate high above
in the axis of the torus, are detected. Their X-ray spectra is highly
absorbed at low energies implying that the torus has a neutral column
density in the range of $10^{22}$-$10^{24}$ cm$^{-2}$. The environment
around central engine of AGN is expected to be fairly complex. For
example, partially ionized, optical thin gas along our line of sight into the nuclei of several AGN can absorb the soft X-rays. This partially ionized material has come to be known as the ``warm absorber'' \citep{Halpern84}. The geometry and the position of these warm absorbers are uncertain. Some AGN show evidence for high-velocity outflows \citep[e.g.][]{Chartas2002, Reeves2003, Dadina2004, Dasgupta2005, Mark2006, Braito2007, Reeves2008, Cap09}. Moreover with the aid of high resolution, narrow band imaging (or integral field spectroscopy) and {\it HST} imaging, it was possible to identify extended regions of highly ionized material around the central engine  also known as `ionization cones' with a size of a few up to 15$-$20 kpc which suggests an anisotropic escape of photons from AGN confined to a cone by a dusty torus \citep{Wilson1993}. Such ionization cones were studied in NGC 1068 and in several other
Seyfert 2 galaxies progressively using dispersive grating spectroscopy
onboard {\it Chandra} and {\it XMM-Newton} \citep[e.g.][]{Sako2000,
Young2001, Kink2002, Ogle2003, Schurch2004, Guai2008, Evans2010,
Dadina2010}. Recent work on Seyfert 2 NGC 4945 based on {\it
XMM-Newton}, {\it Suzaku}, {\it Swift}-BAT and {\it Chandra} data has
revealed detailed characteristics of the circumnuclear reflector. The
study implies the distance of the reflector $\ge$ 35$-$50~pc and the
{\it Chandra} imaging suggests a resolved, flattened, $\sim150$ pc
long clumpy structure, whose spectrum is explained by the cold
reflection of the primary AGN emission \citep{Mari2012}. Constraining
the location and geometry, knowledge of the state and dynamics of
these different components that make up the environment around AGN, is
important to understand how the central engine accretes matter and any
feedback the AGN may have in terms of matter outflow to the galaxy.

An interesting aspect of several Seyfert 2 AGN is the presence of an excess
power law like emission at low energies ($ \sim 1$ keV), which may be
due to scattered emission \citep{Turner1997, Awaki2000}. In this scenario, the direct
low energy photons of the source are absorbed by the torus. There is
an extended scattering medium along the axis of the torus, which
scatters some of the photons to our line of sight. Significant amount
of work has been done using {\it ASCA}, {\it BeppoSAX}, {\it
XMM-Newton} and {\it Suzaku} data, to address the nature of the
scattering medium and its implications towards understanding the
excess soft X-ray emission and geometrical structure in Seyfert 2
AGN. For instance, \cite{Turner1997} studied in detail the importance
of X-ray scattering and reflection for a sample of 25 Seyfert 2 AGN
using {\it ASCA} data. They found a case of Mrk 3 whose spectrum could
be explained by the presence of hot scattering gas in the soft X-ray
regime with scattering fraction cover 0.02\%$-$5\%. A similar study using {\it ASCA} investigated the physical conditions of the scattering material in a sample of scattering-dominated AGN. The study suggested that the soft X-rays
represent scattered light, similar to optical polarized broad lines
\citep{Netzer1998}. \cite{Guai1999} explained the soft X-ray continuum of NGC 1068 and Circinus galaxy observed by {\it
BeppoSAX} due to the
combined presence of a scattered power law emission and an optically
thin plasma emission and estimated column densities of the warm
scatterer as $\lesssim$ 10$^{21}$~cm$^{-2}$ and $\sim$ a few $\times$
10$^{22}$~cm$^{-2}$ respectively. \cite{Awaki2000} studied a sample of
six Seyfert 2 galaxies with optical polarized broad lines using {\it
ASCA} data. The work revealed significant soft X-ray emission owing to
the scattered light from their nuclear emission. They further
estimated a few percent larger scattering efficiency than found for
Seyfert 2 AGN without optical polarized broad lines. \citet{Pappa2001} modeled the scattering emission as a secondary power law component in the {\it ASCA} spectra of a sample of eight Seyfert 2 galaxies and a warm scattering medium in NGC 4151 was
observed by \cite{War2001}.

{\it Suzaku} broadband observations of obscured Seyfert galaxy MCG~$-5-23-16$ revealed a soft excess below 1 keV which was explained by emission from scattered continuum photons and distant photoionized gas. It was deduced that this source was viewed at moderate ($\sim 50^o$) inclination through Compton-thin matter at the edge of a Compton-thick torus covering $\sim 2\pi$ steradians, consistent with Unified models \citep{Reeves2007}. Further studies with {\it Suzaku} 
showed that for  two sources the scattered light constituted a small
fraction ($\lesssim 0.5\%$) of the nuclear emission \citep{Ueda2007}. This suggested a scenario in which there was a geometrically thick torus with a small
opening angle and/or very small amount of gas responsible for scattering. Another detailed work studied the wide band {\it Suzaku} spectrum of Markarian 3 which was resolved into weak, soft power-law emission, a heavily absorbed power law component, cold reflection, and emission lines. The weak, soft power law emission was considered to be scattered light by ionized gas with a scattering
fraction of $0.9\pm0.2\%$ \citep{Awaki2008}. The possibility of
scattering from an ionized medium was discussed in the {\it Suzaku}
analysis of Seyfert 2 NGC 4945 \citep{Itoh2008}.
A comprehensive study by \cite{Noguchi2010} using {\it XMM-Newton}
observations of 32 obscured AGN investigated their multi-wavelength
properties in relation to the scattering fraction. The sample covered
a broad range of the scattering fraction ($0.1\%-10\%$); with eight AGN
exhibiting low scattering fraction ($\sim0.5\%$) suggesting that they
are buried in a geometrically thick torus with a very small opening
angle. They found no significant correlation between scattering
fraction and far-infrared luminosity and a  weak anti-correlation
between the Eddington ratio and scattering fraction.

The variability of the scattered component as compared to the primary emission
can provide constraints on the geometry of the scatterer. X-ray variability
studies of Mrk 3 demonstrated that unlike the hard X-ray emission, the
soft X-rays did not vary over $\sim$13 year baseline suggesting its
origin from an extended region \citep{Turner1997}.
\cite{Awaki2000} also studied Mrk 3 and found  no variability in the
soft component over a 10 year long period although the hard component
changed by a factor of 6 during 3 years. It was suggested that if most
of the soft X-rays from Mrk 3 were scattered, the large time lag indicated that
the distance of the scatterer from the nucleus is comparable to 6
lt-yr ($\sim2$ pc). This result supported a large scattering region as
also presented by \cite{Turner1997}.

IRAS 18325-5926 is an IRAS selected Seyfert 2 galaxy at a redshift $z
= 0.0198$. It was first observed in X-rays by {\it Ginga}
\citep{Iwa95} which showed that the spectral index of the source
$\Gamma \sim 2.2$ is steeper than normal. The {\it ASCA} observation
of this source revealed a quasi-periodic modulation in $\sim 16$ hours
timescales \citep{Iwa98} which was not seen in later {\it RXTE} or
{\it Beppo-SAX} lightcurves \citep{Iwa04}. It is one of the few AGN
which shows a clear broad Iron line \citep{Iwa96} which is possibly
due to reflection of an highly ionized disk \citep{Iwa04}. {\it
Beppo-SAX} observation of this source hinted at a possible roll-over
at $\sim 30$ keV \citep{Iwa04}. Recently, \cite{Zhang2011} reported
the detection of warm absorbers in this source with {\it Chandra}
HETGS spectra. They found an intrinsic absorbing line system with an
outflow velocity $\sim 400$ kms$^{-1}$ which is contributed by two
warm absorbers with outflowing velocities of 340$\pm$110 kms$^{-1}$
and 460$\pm$220 kms$^{-1}$, respectively. \cite{Mocz11}
also found warm absorbers; and detected the absorption features (in the form of a broad trough) in the
vicinity of the $> 7$~keV iron K edge.

IRAS 18325-5926 also shows a soft component which may be due to a scattered
component \citep{Iwa96}. The scattered emission can only respond to variations in the intrinsic photons on a timescale longer than the light crossing one. It is
fortuitous that the hard ($> 2$ keV) X-ray spectrum of IRAS 18325-5926
shows rapid variations on timescales $\sim 10^4$ s and hence is an
ideal source where the size of the scattering emission can be
constrained. Here, we analyze {\it XMM-Newton} and {\it Suzaku}
observations of the source to investigate the timescale on which the
scattered emission responds to the primary one. The fraction of
scattered photons is related to the total column density of the medium
and since the scattered photons are not absorbed, the scattering
medium should be highly ionized as we discuss in this work. Hence,
these physical constraints allow us to estimate the the size and
density of the scattering medium.

In the next section we describe the {\it XMM-Newton} and {\it Suzaku}
observations and the details of data reduction process. In \S 3, the
time-averaged spectra of the source are modeled and in \S 4 the
spectral variability in different timescales are investigated. In \S
5, the size and density of the scattering medium are estimated and the
results and conclusions are presented in \S 6.

\begin{deluxetable}{ccc}
  \tabletypesize{\scriptsize} \tablewidth{0pt}
  \tablecaption{Observation details of IRAS 18325-5926 }
  \tablehead{\colhead{Camera} & \colhead{Exposure Time(ks)} &
    \colhead{Count Rate (s$^{-1}$)} } \startdata & ObsID: 0022940101
  ({\it XMM-Newton}) \\ \colrule
  \ep{} & $3.018$  & $1.36$  \\ \\
  MOS1  & $62.820$  & $1.08$ \\ \\
  MOS2 & $62.880$ & $1.07$\\ \colrule & ObsID: 0022940201 ({\it
    XMM-Newton}) \\ \colrule
  MOS1 & $50.670$ & $1.14$  \\ \\
  MOS2 & $50.690$ & $1.14$ \\ \colrule & ObsID: 702118010 ({\it
    Suzaku}) \\ \colrule
  PIN   & $69.193$ & $0.08$ \\ \\
  XIS 0 & $78.435$ & $1.04$ \\ \\
  XIS 1 & $78.443$ & $1.22$ \\ \\
  XIS 3 & $78.443$ & $1.17$
  \enddata
  \label{Obs_det}
\end{deluxetable}

\section{Observation and Data Reduction}
Two observations from \xmm{} during 2001 March 5 (ObsID: 0022940101)
and March 6 (ObsID: 0022940201) and one observation from {\it Suzaku}
2007 October 10 (ObsID: 702118010) have been analyzed in this work. Details of the observations are shown in Table \ref{Obs_det}.
 
\subsection{XMM-Newton}
The {\it EPIC pn} and {\it MOS} cameras were operated in the small
window mode using medium filter. Data was reduced using version 12.0 of
the Scientific Analysis software (SAS). The exposure time for {\it
  EPIC-pn} camera was too small for first observation (ObsID:
0022940101) and the camera was not in the working mode for the second observation
(ObsID: 0022940201), therefore analysis were done only with the two
{\it EPIC-MOS} cameras. The correction for background flaring is done
by setting a threshold count level of 1.5 counts $s^{-1}$ for the first
observation. For the second {\it XMM-Newton} observation,the background flares were excluded by choosing appropriate time filters which resulted in $\sim 53$ ks data for each {\it MOS} camera. For spectral analysis, the data was filtered to the good X-ray events (FLAG==0) with
pattern $\le$ 12 in the energy range 0.2-15 keV. To extract the source
spectra a circular region of radius 40$\arcsec$ for both {\it MOS1}
and {\it MOS2} centered at the peak position of IRAS 18325-5926 were
used. The background spectra was extracted from a nearby circular
region, free of sources. The source spectra were re-binned with a
minimum of 100 counts per channel.  The re-binned data were analyzed
using {\it {XSPEC}} in the energy range $0.3-10.0\kev$.  For both
observations, it was verified that the {\it MOS1} and {\it MOS2}
spectra were consistent with each other and hence all spectral
fittings were done on the combined data.

\subsection{Suzaku}
Of the four XIS instruments only three of them were in the operating
mode. So observations were taken from three XIS (XIS0, XIS1 and XIS3)
and the HXD (PIN) sensors. The XIS was in normal clocking mode. {\it
  Suzaku} {\it {FTOOL XISRMFGEN}} and {\it { XISSIMARFGEN}} is used to
generate response matrices and ancillary response files
respectively. For the extraction of XIS spectra a circular region of
$4.2'$ radius centered on the source were used, while backgrounds were
extracted from the outer annulus region free of sources. We use an appropriate
scaling factor for the background to take into account the different extraction areas of the
source and the background.
\begin{figure}[htp]
  \begin{center}
    \includegraphics[height=1.02\linewidth, angle=-90]{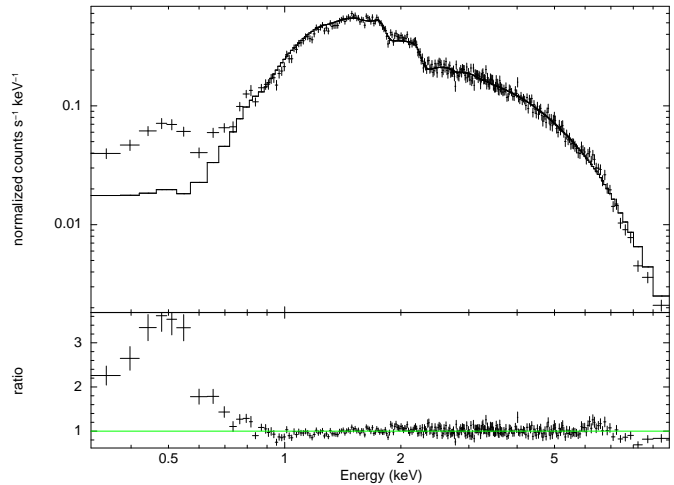}
  \end{center}
  \caption{The \xmm{} 0.3-10.0\kev spectra of IRAS 18325-5926 from the
    combined MOS spectra for the ObsID: 0022940101. The solid line
    indicates the 1-10\kev \ best fitting absorbed power law model
    (upper panel). A weak excess emission can be seen below 1\kev \ in
    the count rate ratio plot (lower panel)}
  \label{soft_exc}
\end{figure}

The source spectrum was re-binned with a minimum of 300 counts per
channel for all the three XIS sensors and 60 counts per channel for
HXD(PIN). The {\it {XSPEC}} package was used for spectral
analysis. The energy range was set to $0.6-10.0\kev$ for XIS and
$12-50\kev$ for HXD. The normalization factor between XIS and HID
spectra was fixed to 1.18.  
A sharp absorption line around $1.8\kev$
was seen in the 0.6-10 keV energy band. It was found that this line
feature is not present in all the XIS instruments. Close examination of
the line reveals that the line is at the silicon K-edge and the
feature could be due to incorrect response calculation. So the energy
range $1.79\ -\ 1.89\kev$ is not considered for spectral
fitting. Apart from this feature it was verified that all the XIS
spectra were consistent with each other. So all the XIS spectra were
combined for the analysis.
\section{Spectral Analysis}

\begin{figure}
\includegraphics[height=1.02\linewidth, angle=270]{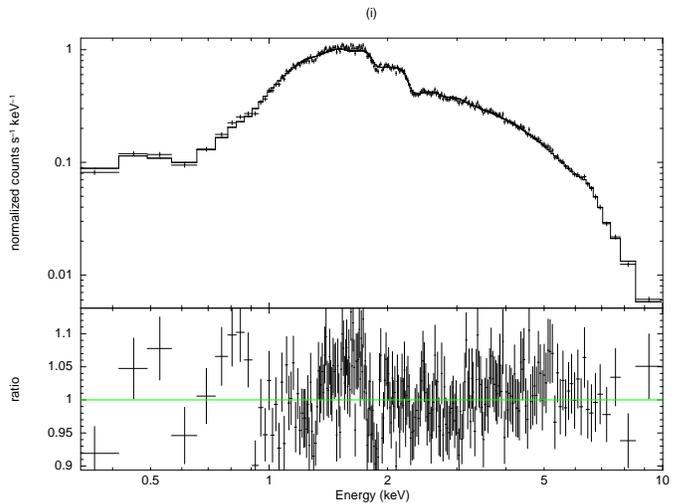}
\caption{ The spectra, best-fit model and deviations of the observed data from the model are plotted for the first {\it
  XMM-Newton} observation. The solid line represents the model based on best-fit parameters listed in Table \ref{Spec_par}. The figure has been rebinned for display purpose only.}
  \label{spec_xmm_suz}
\end{figure}

\begin{figure}
  \includegraphics[height=0.9\linewidth,angle=0]{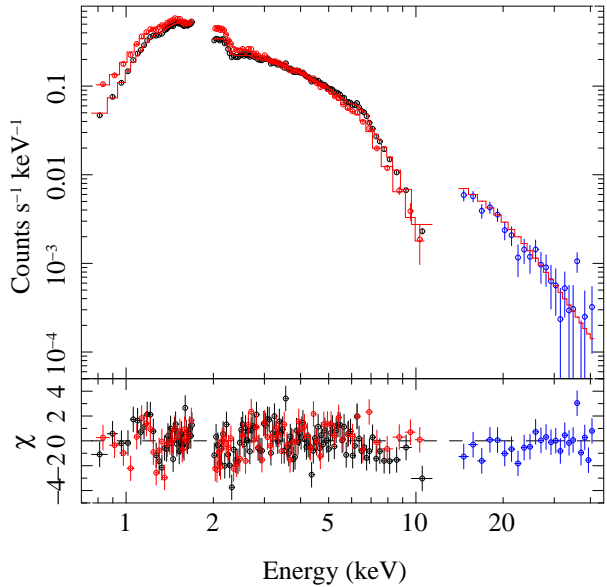}
  \caption{ The Suzaku spectral, best-fit model and deviations of the
   observed data from the model. The figure has been rebinned for display purpose.}
 \label{spec_xmm_suz2}
\end{figure}
To understand the spectra of the source, All three data sets were
initially fitted by an an absorbed power law model.  Galactic
absorption was fixed to column density of $6.47 \times 10^{20}$
cm$^{-2}$. To this an additional neutral absorption at the redshift
of source was included. The first {\it XMM-Newton} data with best fit
model and their ratio are plotted in Fig \ref{soft_exc}. The ratio
plot clearly shows two excesses. One at low energies $\sim 0.5\kev$
which could be due to a scattered component and other at $\sim 6\kev$
due to the K$\alpha$ Iron line emission.

The Iron line emission feature in both the {\it XMM-Newton} and the {\it Suzaku} 
observations can be fitted by a broad Gaussian feature. The centroid energy is at $\sim 6.6\kev$ and is
broad with $\sigma \sim 0.3\kev$ which is consistent with the {\it ASCA} 
observations \citep{Iwa96}. The width of the line suggests a
relativistic origin for the line. The relativistic Iron line model in
{\it xspec}, ``diskline'' can also represent the feature. However, the
line is too weak to constrain most of the parameters.  For example,
the first observation of {\it XMM-Newton}, provides a line energy of
$6.44_{-0.09}^{+0.08}\kev$ and an inclination angle of $41^{+6}_{-7}$
degrees, if the other parameters (i.e. the emissivity index $q = -2$,
inner $R_{in} = 6$ and outer radius $R_{out} = 1000$) are fixed at
standard values. The line energy is less than what is expected from a
highly ionized disk ($\sim 6.9\kev$) as modeled by \cite{Iwa04}.
However, the present data is not good enough to undertake a detailed
analysis of the complex iron line feature.  Further complications are
evident by the presence of an absorption edge at $\sim 8.2\kev$ in the
second {\it XMM-Newton} observation. Thus, in this work we model the
Iron line feature using the ``diskline'' model and concentrate instead
on the low energy excess.

The low energy excess is a broad feature and cannot be due to presence
of narrow emission lines. We have checked the RGS data and confirmed
that there are no strong narrow emission lines in the soft band as also revealed in the \chandra~ observation \citep{Mocz11}. Instead, it can be modeled as an additional unabsorbed
(i.e. only affected by the Galactic absorption) power law with the
spectral index tied to that of the intrinsic one. The normalization of this additional
power law is the only free parameter in the fitting \citep{Trippe2010}. The
best fit spectral parameters are tabulated in Table \ref{Spec_par} and
the spectra with best-fit model and residuals are shown in Fig. \ref{spec_xmm_suz} and Fig. \ref{spec_xmm_suz2} for the
first {\it XMM-Newton} and {\it Suzaku} observations respectively. The spectral
fits are reasonable with no obvious systematic residuals.

\begin{deluxetable}{lccc}[htp]
  \tabletypesize{\scriptsize} \tablewidth{0pt} \tablecaption{Best-fit
    spectral parameters for the three observations. I:{\it XMM-Newton
    } (0022940101), II: {\it XMM-Newton } (0022940201) III: {\it
      Suzaku} (702118010)} \tablehead{\colhead{Model Parameters} &
    \colhead{I} & \colhead{II} & \colhead{III}} \startdata
  & & & \\
  $N_H$\tablenotemark{a}  & $1.08\pm0.02$ & $1.06\pm0.02$ & $1.14\pm0.02$ \\
  $E_{Edge}$\tablenotemark{b} (\kev) & -- & $8.17_{-0.33}^{+0.25}$ &--\\
  $\tau$  & - & $0.26_{-0.10}^{+0.11}$ &--\\ \\
  $E_{Line}$ \tablenotemark{c} (\kev) & $6.44_{-0.09}^{+0.08}$ & $6.47_{-0.08}^{+0.07}$ & $6.63\pm0.05$\\
  $i (deg)$ & $41_{-7}^{+6}$ & $48_{-8}^{+18}$ & $50$(f) \\
  $r_{in}$ ($r_g$) & $6$ (f) & $6$(f) & $110.8_{-33.5}^{+49.6}$ \\
  $n_D$ & $3.4\pm0.7\times10^{-5}$ & $3.8_{-0.9}^{+1.0}\times10^{-5}$ & $3_{-0.5}^{+0.6}\times10^{-5}$ \\ \\
  $\Gamma$ \tablenotemark{d}   & $1.92\pm0.02$ & $1.97\pm0.02$ & $2.14\pm0.015$ \\
  $n_{PL_P}$ & $5.37\pm0.14$ & $5.93\pm{0.17}$ & $13.5\pm0.3$ \\ \\
  $n_{PL_S}$ \tablenotemark{e} & $0.18\pm0.01$ & $0.17\pm0.01$  & $0.28\pm0.04$ \\ \\
  $\chi^2/dof$ & $888.85/796$ & $738.33/694$  & $1045.6/859$\\
  \enddata
  \tablenotetext{a}{Absorption column density in units of $10^{22}$
    cm$^{-2}$. The Galactic absorption of $6.47 \times 10^{20}$
    cm$^{-2}$ has been included in all spectra.  $^b$ Absorption edge
    component with energy $E_{Edge}$ and optical depth $\tau$ required
    only for observation II. $^c$Diskline model where $E_{line}$ is
    the rest frame energy, $i$ is the inclination angle and $n_D$ is
    the normalization of the model in $10^{-5}$ photons
    cm$^{-2}$s$^{-1}$.  The emissivity index $q = -2$, inner radius
    $R_{in} = 6$ and outer radius $R_{out} = 1000$ have been fixed to
    standard values. $^d$The primary power law with photon index
    $\Gamma$ and normalization, $n_{PL_P}$ in units of $10^{-3}$
    photons cm$^{-2}$s$^{-1}$. $^e$Scattered power law emission with
    photon index same as the primary one and normalization, $n_{PL_S}$
    in units of $10^{-3}$ photons cm$^{-2}$s$^{-1}$.}
  \label{Spec_par}
\end{deluxetable}

\section{Variability Studies}

As expected, Table \ref{Spec_par} shows that the spectrum of IRAS
18325-5926 was different in the \suzaku observation of October 2007
as compared to the \xmm ones of March 2003. In particular, the
normalization of the primary power law changed by nearly a factor of
two. The normalization of the scattered power law emission also reveal
a significant increase by $\sim 50$\% which shows that on timescales
of years the scattered emission is variable, most probably in response
to the primary continuum. The ratio of the scattered to the primary
continuum also changed from $\sim 0.03$ during 2003 to $\sim 0.02$ in
2007, suggesting that the scattering medium itself is variable on such
long timescales.

The two \xmm observations are about a day apart and Table
\ref{Spec_par} reveals that the normalization of the primary power law
does vary on this timescale, by increasing from $5.37\pm0.14\times 10^{-3}$ to
$5.93\pm0.17\times 10^{-3}$ photons cm$^{-2}$ s$^{-1}$. However the
normalization of the scattered component does not respond to this
variation on $\sim 10^5$ s, remaining nearly constant at $0.18\times
10^{-3}$ photons cm$^{-2}$ s$^{-1}$.

\begin{figure}
  \includegraphics*[height=0.78\linewidth,angle=0]{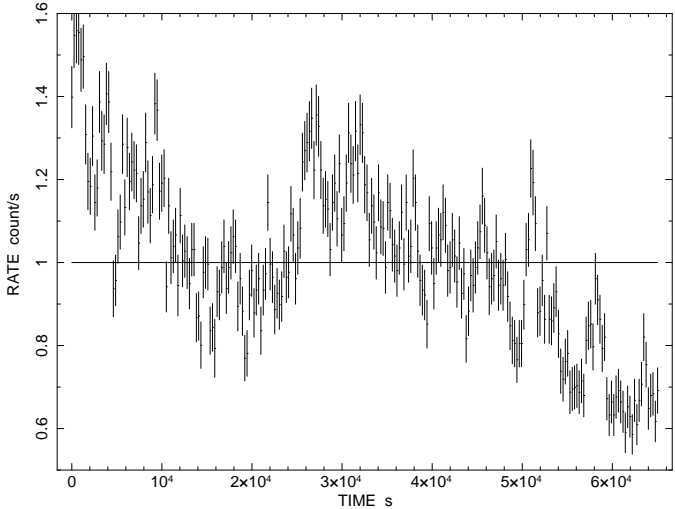}
  \caption{ X-ray lightcurve of IRAS 18325-5926 from the \xmm{}
    observation in the energy range $0.3$-$10$\kev~for ObsID
    0022940101. The solid horizontal line at $\sim 1$ count/s is used
    to divide the lightcurve into low and high flux states.  }
  \label{xmmlightcurves}
\end{figure}

The source shows significant variability on a shorter timescale of
$10^4$ s, as is evident in the lightcurve of one of the \xmm
observations shown in Fig \ref{xmmlightcurves}.  
We investigate whether the scattered continuum is variable on this timescale by
dividing the data into two states defined as high ( count rate > 1
counts/s) and low (count rate < 1 counts/s) flux levels. Spectral
fitting of each flux level for the first \xmm observation, reveals
that the primary power law normalization varied from $5.05 \pm 0.04$
to $7.03\pm 0.02 \times 10^{-3}$ photons cm$^{-2}$s$^{-1}$. However
the scattered component normalization did not seem to respond to this
variation with $1.87 \pm 0.1$ (low flux level) to $1.98 \pm 0.1 \times
10^{-4}$ photons cm$^{-2}$s$^{-1}$. A similar result was obtained for
the second \xmm observation where the primary component normalization
changed from $5.36 \pm 0.08$ to $7.43\pm 0.1 \times 10^{-3}$, but
there was no detectable change in the scattered one with $1.80 \pm
0.1$ (low flux level) to $1.75 \pm 0.1 \times 10^{-4}$ photons
cm$^{-2}$s$^{-1}$ (high flux level). 

\begin{figure}
  \includegraphics[width=1.\linewidth]{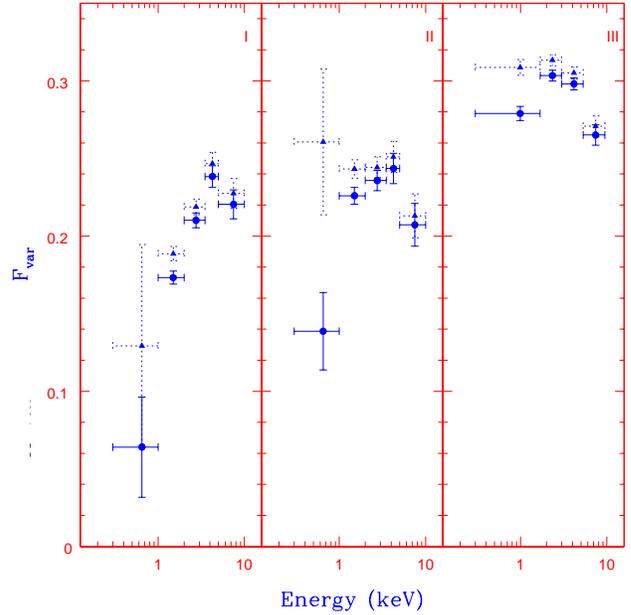}
  \caption{The fractional variation, $F_{var}$ (filled circles) as a
    function of energy for (I) ObsID: 0022940101, (II) 0022940201 of
    \xmm{} and (III) XIS spectrum of {\it Suzaku}. Also plotted is the
    fractional variation that the primary component should have (filled triangles with dotted errorbars), $F_{P,var}$, in the absence of any variation in the scattered
    one.}
  \label{rms}
\end{figure}

Further evidence that the
scattered component does not respond to the primary one comes from
analyzing the variability as a function of energy. For this, we extracted lightcurves for XMM-Newton observations in the energy bands 0.3-1.0 keV, 1.0-2.0 keV, 2.0-3.5 keV, 3.5-5.0 keV and 5.0-10.0 keV, respectively and for Suzaku observation, 0.3-1.68 keV, 1.68-3.0 keV, 3.0-5.32 keV and 5.32-9.5 keV, respectively. For extracting these lightcurves we have selected the good time intervals excluding the period of high background flares. In Fig \ref{rms},
the fractional root mean square variability amplitude ($F_{var}$) is
plotted against energy for all three data sets. $F_{var}$ is the
square root of the excess variance which is the measured variance
minus the expected variance due to Poisson noise. The measurement
error on $F_{var}$ is computed using the formalism given by
\cite{Vau03}. For all three data sets there is a sharp decrease in the
variability below $1\kev$ where the scattered component is a
significant fraction of the total spectrum. This is indicative of a
variable primary and a constant scattered component. One can estimate
the variability that must be present in the primary component,
$F_{P,var}$ to reproduce the observed $F_{var}$, as
\begin{equation}
  F_{P,var} = F_{var}\;\ \left[
    \dfrac{c_T}{c_{P}}\right] 
\end{equation} 
where $c_T$ is the total count rate detected in the energy bin and
$c_P$ is the count rate in that bin due to the primary component
alone. Fig. \ref{rms} shows $F_{P,var}$ as a function of energy
(triangles with dotted lines as error bars). Thus, a nearly energy
independent variability of the primary component, $F_{P,var}$ with a
non-varying scattered component can explain the total observed
variability of the source.

Thus the variability of the source indicates that while the primary
component varies in a wide range of timescales, the scattered
component responds only on timescales longer than a day ($10^5$ s).

\section{Geometry Of the Scatterer}

Since the scattered component does not respond to the continuum on
timescales of $10^5$ s, the scattering medium must be larger than $ >
3 \times 10^{15}$ cm. A simple geometry for the medium is a funnel
shaped volume along the axis of the accretion disk.  If the opening
angle of the cone is $\alpha \sim 45^o$, the fraction of photons that
will be scattered is $\eta \sim (1 - \hbox{cos} \alpha) \tau$, where
$\tau = N_H \sigma_T$ is the Thomson optical depth of the
medium. Thus, we can estimate the column density $N_H$ as
\begin{equation}
  N_H \  \sim \ 1.5 \times 10^{23}\ \hbox{cm}^{-2} \ \left(\frac{1 - \hbox{cos}\alpha}{0.3}\right)^{-1} \ \left(\frac{\eta}{0.03}\right)
\end{equation} 
where $\eta \sim 0.03$ is the observed ratio of the normalizations of
the primary to the scattered power law.

Such a high column density would absorb the soft X-ray photons from
the source (instead of scattering them) unless the medium is highly
ionized. For that to happen, the ionization parameter $\xi = L/nR^2$
must be larger than a critical value $\xi_c > 10^3$. This naturally
puts an upper limit on the size of the region $R < L/\xi_c N_H$ or
\begin{eqnarray}
  R \ &  <  & \ 3 \times 10^{17} \ \ \ \hbox{cm}\ \left(\frac{\xi_c}{10^3}\right)^{-1} \ \left(\frac{\eta}{0.03}\right)^{-1} \nonumber \\
  \ &  & \ \times \left(\frac {L}{ 5 \times 10^{43} \ \hbox{ergs s}^{-1}}\right)\ \left(\frac{1 - \hbox{cos}\alpha}{0.3}\right)
\end{eqnarray}
The X-ray luminosity of the source in the $0.3$-$10\kev $ band is $L_X
\sim 1.5 \times 10^{43}$ ergs s$^{-1}$ for the \xmm observations assuming $H_{\rm 0}$=70 km s$^{-1}$ Mpc$^{-1}$. The ionizing luminosity $L$ which includes photons with energy $> 13.6$ eV
would be about a factor of few higher and is taken as $\sim 5 \times
10^{43}$ ergs s$^{-1}$.  Thus the scattering medium can be constrained
to have a column density $N_H \sim 1.5 \times 10^{23}$ cm$^{-2}$, a
size in the range $3 \times 10^{15} \ < \ R \ < 3 \times 10^{17}$ cm
and an average number density $n \sim 5 \times 10^{6}$ cm$^{-3}$.

In this interpretation, the scattering medium is highly ionized and
hence if the source was viewed in a lower inclination angle (like in a
Seyfert 1), the spectrum will not be absorbed. However, a small
fraction of the gas, (with column density $\simless 10^{22}$
cm$^{-2}$) which is furthest away from the source could be partially
ionized. This partial ionized gas could be the origin of the warm
absorbers observed in many Seyfert 1.  This would imply that the warm
absorber clouds are at a distance of $> 10^{17}$ cm from the black
hole.  This is interesting since spectral fitting of the warm
absorbers cannot constrain this distance and there always has been a
large uncertainty regarding their location.

The present analysis does not provide information about the dynamic
nature of the scattering medium as to whether it is an outflow or
inflow. If we assume the medium to be outflowing in typical speeds of
$v \sim 1000$ km s$^{-1}$, the mass outflow rate can be estimated to
be
\begin{eqnarray}
  \dot M_{out} \  & \sim &  \ \ \ m_p \ n \  v \ (4 \pi R^2) \ (1 - \hbox{cos}\alpha) \ \   \sim \ \ 3 \ \times \ 10^{23} \ \hbox{g/s}   \nonumber \\
  &   &   \times \left(\frac{n}{5 \times 10^6 \hbox{cm}^{-3}}\right) \left(\frac{v}{10^8 \hbox {cm s}^{-1}}\right) \left(\frac{R}{10^{16} \hbox {cm}}\right)^2  
\end{eqnarray}
The mass accretion rate in the accretion disk can be estimated from
$L_{bol} \sim \eta \dot M_A c^2$ as being $\sim 10^{24}$ g/s for a
radiative efficiency, $\eta \sim 0.1$ and bolometric luminosity $L_{bol}
\sim 10^{44}$ ergs s$^{-1}$, which is consistent with the results obtained by {\it Chandra} analysis of this source \citep{Mocz11}. Despite the large uncertainties in these
estimations, it seems that the scattering medium would require a mass
outflow rate which is a significant fraction of the accretion one. The
kinetic power carried by the outflow $\dot E \sim \dot M_{out} v^2
\sim 3 \times 10^{39}$ ergs s$^{-1}$ is small compared to the
radiative luminosity. However, these estimates depend upon the information on velocity which is unknown.

An alternate interpretation to the scattering medium described above,
is that the secondary component is reflection from a highly ionized
inner layer of the torus surrounding the black hole.  From the lack of
variability, the radius of the torus, $R_t \sim 3 \times 10^{16}$ cm and
for it to be highly ionized would require, the density $n < L/\xi_c
R_t^2$. For the torus to efficiently reflect the X-ray radiation, it
must be optically thick to scattering i.e. its column density $N_{t} >
10^{24}$ cm$^{-2}$. This implies that the width of the torus $\Delta
R_t/R_t = N_t \xi R_t/L \gtrsim 1$, making the geometry unlikely. Hence
we prefer the scattering geometry above as a more viable scenario.

\section{Results and Conclusions}
The spectrum of IRAS 18325-5926 is well modeled by an absorbed power law and a secondary power law component to describe the low energy excess feature $\lesssim 0.5\kev$ in the form of a scattered emission. We found that the scattered component varies on timescales longer than $10^{5}$ s unlike the primary emission which varies on a wide range of timescales ($10^{4} - 10^{8}$ s). This indicates the size of the scattering medium to be greater than $10^{16}$ cm. The observed ratio of the scattered to primary flux ($\sim  0.03$) gives the column density for the scattering medium to be $\sim$ 10$^{23}$ cm$^{-2}$. In order to facilitate scattering instead of absorption, the location of such a scattering medium should be less than $10^{17}$ cm from the X-ray source for it to be highly ionized. So we are also able to localize the position and extent of scattering region to be $\sim$ a few times $10^{16}$ cm, with an average particle density of $\sim 10^{6}$ cm$^{-3}$. Hence, the location of the torus is $\gtrsim$ 10$^{16}$ cm. Besides, the spectral analysis of two XMM-Newton and a Suzaku observation also reveal the presence of a broad Iron line feature in the source. For Seyfert 2 AGN, with column density $N_{H}$ > 10$^{24}$ cm$^{-2}$, the matter is optically thick to Compton scattering which makes the nucleus almost invisible. Hence the broad iron line cannot be detected in this scenario. However, for Seyfert 2 AGN with relatively low values of $N_{H} \sim$ 10$^{23}$ cm$^{-2}$, where one sees through the outer edge of the putative cold torus, central X-ray emission above a few keV can penetrate the torus making the nuclear source visible to the observer. This picture would allow the detection of broad Fe K emission line consistent with the unification scheme. The presence of absorption edge at $\sim$ 8.2 keV in the second XMM-Newton observation of IRAS 18325-5926 (ObsId:0022940201) may suggest the possible link to the scattering medium, but it also depends on whether the X-rays are absorbed by the scattering medium or not.

A speculative but consistent scenario for the AGN seems to be that the
source has a bi-polar outflow with a mass outflow rate comparable to
the accretion one. However with a kinetic power much smaller than the radiative luminosity. The outflow is highly ionized to a distance of $\sim 10^{17}$ cm, with a column density of $N_H \sim 10^{23}$
cm$^{-2}$, beyond which there may be partially ionized gas.  The
source is surrounded by a torus at a distance of $\sim 10^{16}$
cm. A fraction of the primary X-rays scatter into our line of sight from the
ionized outflow. A more systematic study of several such sources and
theoretical hydro-dynamical simulations have to be undertaken to
validate this geometry.
\section{Acknowledgements}
This work is based on observations obtained with {\it XMM-Newton}, an ESA science mission with instruments and contributions directly funded by ESA Member States and NASA. This research has made use of data obtained from the {\it Suzaku} satellite, a collaborative mission between the space agencies of Japan (JAXA) and the USA (NASA). We would like to thank the {\it XMM-Newton} and {\it Suzaku} mission teams for developing excellent instruments. We thank the anonymous referee for the careful reading of the manuscript and useful comments.

\end{document}